\documentclass[]{aa}
\usepackage{graphicx}
\usepackage{here}
\usepackage{color}
\usepackage{natbib}

\newcommand{\nc}{\newcommand}
\nc{\bec}{\begin{center}}
\nc{\enc}{\end{center}}
\nc{\beq}{\begin{equation}}
\nc{\enq}{\end{equation}}
\nc{\bei}{\begin{itemize}}
\nc{\eni}{\end{itemize}}
\nc{\bee}{\begin{enumerate}}
\nc{\ene}{\end{enumerate}}
\nc{\namely}{{\it viz.}}
\nc{\td}{T$_{\rm d}$}
\def\lsun{$L_\odot${}}
\def\micron{\hbox{$\mu$m}}

\def\dep100{$\tau_{\rm 100}$}
\def\dep150{$\tau_{\rm 150}$}
\def\12CO{$^{12}$CO}
\def\13CO{$^{13}$CO}
\def\cplus{C$^{+}$}
\def\ozero{O$^{0}$}
\def\fps100{{\bf{FPS100}}}
\nc{\ea}{et al.}

\begin{document}


\title{Mapping of Large Scale $158$~\micron\ [CII] Line Emission: Orion A}

\author{ B. Mookerjea \inst{1,4,5}, S.K. Ghosh\inst{1}, H. Kaneda\inst{2}, T. Nakagawa\inst{2}, D.K. Ojha\inst{1},\\T.N. Rengarajan\inst{1,6}, H. Shibai\inst{3} \& R.P. Verma\inst{1} }

\offprints{B. Mookerjea}

\institute{Tata Institute of Fundamental Research,
            Homi Bhabha Road,
            Mumbai (Bombay) 400 005, India
\and
            Institute of Space and Astronautical Science,
	    Kanagawa 229, Japan
\and
            Dept. of Physics, Nagoya University, Nagoya 464, Japan
\and
            Joint Astronomy Programme, Dept. of Physics, Indian
	    Institute of Science, Bangalore 560 012, India
\and
            I. Physikalisches Institut, 
            Universit\"{a}t zu K\"{o}ln, 
	    Z\"{u}lpicher Strasse 77, 
	    D-50937 K\"{o}ln, Germany
\and
            Instituto Nacional de Astrofisica, 
            Optica y Electronica, 
	    Tonantzintla, Puebla 72840, México, Mexico}
\date{}
\titlerunning{$158$~\micron\ [CII] emission from Orion A}
\authorrunning{B. Mookerjea et al.}


\abstract{

We present the first results of an observational programme undertaken to
map the fine structure line emission of singly ionized carbon ([CII]
$157.7409$~\micron) over extended regions using a Fabry Perot
spectrometer newly installed at the focal plane of a $100$~cm
balloon-borne far-infrared telescope. This new combination of
instruments has a velocity resolution of $\sim 200$~km/s and an angular
resolution of $1$\farcm $5$.  During the first flight, an area of
$30$\arcmin $\times 15$\arcmin\  in Orion~A was mapped. These
observations extend over a larger area than previous observations, the
map is fully sampled and the spectral scanning method used enables
reliable estimation of the continuum emission at frequencies adjacent to
the [CII] line. The total
[CII] line luminosity, calculated by considering up to 20\% of the
maximum line intensity is $0.04$\% of the luminosity of the far-infrared
continuum. We have compared the [CII] intensity distribution with the
velocity-integrated intensity distributions of \13CO(1-0), CI(1-0) and
CO(3-2) from the literature.  Comparison of the [CII], [CI] and the
radio continuum intensity distributions indicates that the largescale
[CII] emission originates mainly from the neutral gas, except at the
position of M43, where no [CI] emission corresponding to the [CII]
emission is seen. Substantial part of the [CII] emission from here
originates from the ionized gas.  

The observed line intensities and ratios have been analyzed using the
PDR models by \citet{kaufman99} to derive the incident UV flux and
volume density at a few selected positions. The models reproduce the
observations reasonably well at most positions excepting the [CII] peak
(which coincides with the position of $\theta^{1}$ Ori~C). Possible
reason for the failure could be the simplifying assumption of a
homogeneous plane parallel slab in place of a more complicated
geometry.

\keywords{infrared: ISM -- ISM: lines and bands -- ISM: individual (Orion A)}
}

\maketitle



\section{Introduction}

In recent years it has been established that molecular clouds show definite
evidence of internal structures in the forms of clumps and filaments. The
UV radiation, from external or embedded OB stars, incident on the molecular
clumps dissociates molecules and ionizes atoms within a shell of A$_{\rm v}
\approx 5^{m}$ on the surface of the clumps and forms the Photon Dominated
Regions (PDR). The externally heated PDR gas cools via characteristic FIR
and sub-mm lines of various atomic and molecular species. The fine
structure lines of [CII] $158$~\micron\ and [OI] $63$~\micron\ typically
account for $1$\% of the cooling of the PDRs. One dimensional theoretical
models for PDRs \citep{tielens85,sternberg89}  can explain the observed
intensities of \cplus, C$^{0}$ and \ozero\ lines  but
fail to explain the extended nature of the [CII] emission. However a more
realistic model suggests that the molecular clouds are clumpy in nature
which allows the UV radiation to
penetrate deeper and hence the observed emission consists of the
contribution from many photodissociated clump surfaces along the line of
sight \citep{koster94}. The observed [CII] as well as the CO line
intensities scale with the density of the PDRs and the intensity of the UV
radiation. Most of the high spatial resolution observational studies of
[CII] line emission have concentrated on regions of high density and high
UV radiation intensity. However large scale surveys with COBE
\citep{wright91}, Balloon-borne Infra Red Telescope (BIRT) \citep{shibai91}
and Balloon-borne Infrared Carbon Explorer (BICE) \citep{nakagawa98} have
shown that the diffuse [CII] emission from the lower density regions
contributes significantly to the cooling of the gas in our Galaxy and there
is a significant drop in [CII] to continuum ratio towards regions of active
star formation. Thus large scale high angular resolution studies of
extended [CII] and CO emission from both the active star forming regions
and PDRs with low densities and low UV radiation intensity are needed to
understand the dominant cooling mechanisms under different physical
conditions within the molecular clouds. However the [CII] fine structure
line is observable only from above the earth's atmosphere
using aircraft and balloons, thus restricting the scope of the
observational studies. Here we present the first results of a balloon-borne
programme for observation of the [CII] emission from Galactic star forming
regions.

Situated at a distance of 450 pc, the Orion molecular cloud is the
nearest region of active star formation. It also shows clear evidence of
ubiquity of PDRs \citep{genzel89}. Owing to its proximity and
brightness, numerous observational studies have been directed towards
Orion and these have resulted in improving our understanding of the
interplay between stars and the interstellar matter (ISM).  In addition,
Orion is also a good benchmark source to test any new observing system.
The Orion molecular cloud also has the advantage of being like a ridge
extended along the north-south direction with much smaller extent along
the east-west. This is particularly useful for chopped observational
studies, so that the emission lost due to self-chopping is negligible. 

Previous [CII] observations of Orion were carried out from the Kuiper
Airborne Observatory. These include the mapping of a $6$\farcm $5 \times
10$\arcmin\ region in OMC~1 by \citet{stacey93}, the mapping of
$7$\arcmin $\times 18$\arcmin\ region in the Orion Molecular Ridge
(OMC~1 and 2) by \citet{herrmann97} and heterodyne spectroscopy of M42
by \citet{boreiko88,boreiko96}. 

Here we present a large scale study of [CII] $158$~\micron\ fine-structure
line emission towards Orion A. The map is fully sampled and covers an area
of $30$\arcmin $\times 15$\arcmin. The rest of the paper is organized as
follows: A brief description of the instrument, observational procedure and
parameters are presented in Sect.~2. Sect.~3 presents the data analysis
methods, Sect.~4 presents the results of observation, Sect.~5 presents
comparison with other observations and Sect.~6 presents a discussion of the
intensity profiles and interpretation of the line intensity ratios using a
PDR model.

\section{Instrument \& Observation}

\subsection{FPS100}

A Fabry Perot Spectrometer (FPS) tuned to the [CII] $157.7409$~\micron\
line was installed at the focal plane of the $100$~cm TIFR balloon-borne
FIR telescope \citep{ghosh88}. We refer to this new combination of
instruments as \fps100. The FIR telescope has an f/8 Cassegrain
configuration in which the secondary mirror can be wobbled with an
amplitude of $4$\farcm$5$\ and a frequency of $10$~Hz. The FPS was designed
at ISAS, Japan and in an earlier configuration was used at the focal plane
of the BICE \citep{nakagawa98}. Interfacing of the FPS to the TIFR
telescope necessitated modification of the optical components in order to
achieve f-number compatibility between the two. New electronics was also
designed and  built to implement the observing modes (discussed in
Sect.~\ref{obsmode}). We point out that the FPS on the BICE had an
angular resolution of $15$\arcmin, while on the TIFR telescope the achieved
angular resolution is $1$\farcm $5$. The FPS, in brief is a tandem Fabry
Perot Spectrometer, consisting of two Fabry Perot interferometers: one a
high order interferometer with movable plates to scan the wavelength
(Scanning Fabry Perot; SFP) and the other a low-order interferometer with
fixed plates (Fixed Fabry Perot; FFP) which acts as an order sorter for the
SFP. The two interferometers together with other optics of the spectrometer
and the FIR detector, are cooled to $2$~K with liquid helium. The FIR
detector used is a stressed Ge:Ga photoconductor \citep{hiromoto89} with an
NEP of $3.1 \times 10^{-15}$~Watt/$\sqrt{\rm Hz}$.

\subsection{Observation Modes }
\label{obsmode}

The \fps100\ has been designed for two observation modes, \namely, {\em
chopped} and {\em unchopped}. Both modes involve frequency modulation of
the signal achieved by sweeping the SFP up and down at a preselected
scan frequency and over a preselected velocity range around the expected
position of the [CII] line. The spectral scanning, in addition to being
important for background reduction is also useful for estimating the
neighboring continuum reliably. The primary difference between the two
modes of observation lies in the fact that, in the {\em chopped} mode the
sky chopping (effected by wobbling the secondary mirror) is functional,
while in the {\em unchopped} mode there is no sky chopping. In the {\em
chopped} mode the frequency of chopping is $10$~Hz. For both modes it is
possible to carry out spatial raster scans covering a rectangular box
around the target source.

Thus, the {\em chopped} mode involves sky chopping, spatial and spectral
scanning. For meaningful measurements the frequencies of spatial scan
($\nu_{\rm sp}$), sky chopping ($\nu_{\rm ch}$) and spectral scan
($\nu_{\rm wv}$) in this mode should maintain the following relation:
$\nu_{\rm sp}$ $<$$\nu_{\rm wv}$$<$$\nu_{\rm ch}$. The signal detected in
this mode is the difference between the emission in the two beams and is
processed through linear phase band pass filters and further amplified
(with selectable gains). There after the signal is phase sensitively
detected (with online phase adjustment capability) and digitized for the
telemetry. One of the  outputs of the PreAmplifier for the SPectrometer
(PASP) is used as the output signal for the $chopped$ mode. The signal is
sampled at a frequency of $10$~Hz.

In the $unchopped$ mode, since there is no onboard background subtraction,
the integration time over each spectral element is kept small,
i.e., in effect a series of narrow band observations are done. Thus for
this mode the spectral scan frequency is at least $4$ times higher than
that for the $chopped$ mode.  For this mode the relation between the
different frequencies involved is $\nu_{\rm ch}$ $<$$\nu_{\rm
sp}$$<$$\nu_{\rm wv}$. Since there is no modulation of the signal due to
sky chopping ($\nu_{\rm ch} = 0$), the corresponding output from the PASP
is directly digitized using a $16$ bit ADC. The frequency of spectral
modulation being higher in this mode, the signal is sampled at a frequency
of $40$ Hz.

\subsection{Observation Details}

The \fps100\ payload was flown for the first time from the TIFR Balloon
Facility, Hyderabad, in Central India (Latitude = N$17$\fdg $47$,
Longitude = E$78$\fdg $57$) on 1999 November 25. The observations
presented here were carried out in the {\em chopped} mode. The choice of
the {\em chopped} mode over the {\em unchopped} mode was based
on considerations related to the noise level during the flights. For the
{\em chopped} mode the noise level was found to be consistent with the
pre-flight tests in which the flight condition was simulated by closing
the window of the FPS and keeping the rest of the configuration identical
to that during the flight.

Jupiter was observed for absolute flux calibration as well as for
determining the point spread function (PSF). The angular resolution
achieved during these observations was $1$\farcm $5$. A rectangular region
of $30$\arcmin $\times 15$\arcmin\ towards Orion~A was observed in
approximately $50$ minutes. The spatial scan rate was $0$\farcm $32$/s
and the spectral scan frequency was $0.5$~Hz.

\section{Data Analysis}
\label{data}

The entire dataset for Orion A was split into two parts depending on
whether the SFP is scanned from lower to higher frequency (Up) or in the
reverse sense (Down). The data analysis procedure described in the
following text was applied independently to these two datasets.

The signal measured is a combination of the emission from the astronomical
source and the residual (post sky chopping) foreground radiation
contributed by the atmosphere and the instrument. In the present work we
have estimated the foreground emission by considering the signal at the
edges of the spatial scan lines where there is no apparent contribution
from the astronomical source. As expected the residual emission in the  {\em
chopped} mode was found to be negligible.

Each spectral scan was corrected for foreground radiation. Since the
instrumental profile is much wider than the intrinsic astronomical line
profile, the observed spectral profiles are dominated by the instrumental
profile. Since the Lorentzian profile is a good approximation for the
instrumental profile, we have fitted each spectral profile by a
combination of the Lorentzian profile and a linear function of the
wavenumber $\sigma$,
representing the contribution of the continuum emission at these
wavelengths. The fitted function has the following form:

\begin{equation}
f= \frac{C_{1}}{2\pi}\frac{\Gamma}{(\sigma-\mu_{0})^{2} +
(\Gamma/2)^{2}} + C_{2}\sigma + C_{3}
\end{equation}

\begin{figure}[h]
\begin{center}
\includegraphics[angle=0,width=7.9cm,angle=0]{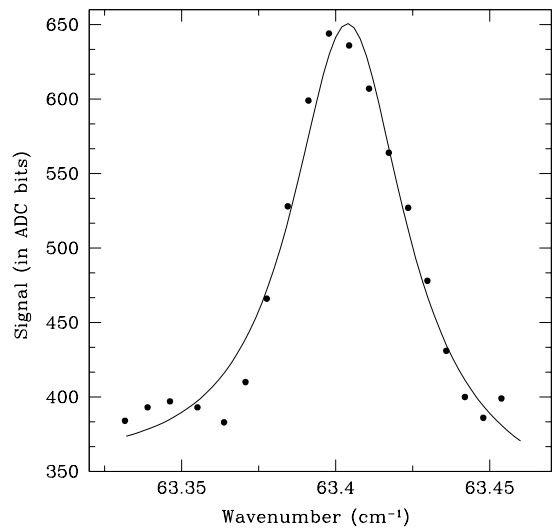}
\caption
{Typical Spectral Profile from Orion Observations. The observed data are
shown as points and the function (combination of a Lorentzian
and a straight line) fitted to the observations is shown as a continuous
line. The fitted function corresponds to $\mu_{0}=63.404$~cm$^{-1}$
and $\Gamma=0.045$~cm$^{-1}$.
\label{specprof}}
\end{center}
\end{figure}

The fitting procedure involves $5$ parameters \namely, C$_{1}$, C$_{2}$,
C$_{3}$, $\Gamma$ and $\mu_{0}$. $\mu_{0}$ and $\Gamma$ are respectively
the central position and FWHM of the line.
 The observed line profile is slightly modified from the
instrumental profile because the astronomical line emission slightly
broadens its width and shifts its central position as well. Thus for
regions with strong [CII] line emission, all the $5$ parameters are kept
free and $\mu_{0}$ and $\Gamma$ are determined accurately. Since the [CII]
linewidth in the region near $\theta^{1}$ Ori C is $5$~km/s
\citep{boreiko88}. Hence the line being not resolved by the {\bf FPS100}
we have kept $\mu_{0}$ and $\Gamma$ fixed for all spectral profiles  and
determined the values of the remaining parameters from the fit.
Fig.~\ref{specprof} presents the observed data (shown as points)
corresponding to a spectral profile and the function (shown as a
continuous line with $\mu_{0}$=$63.404$~cm$^{-1}$ and $\Gamma$=
$0.045$~cm$^{-1}$) fitted to it. This value of $\Gamma$ corresponds to a
velocity resolution of $210$~km/s.

This profile fitting results in two datasets corresponding to the
intensities of the [CII] line and the continuum. Using the telescope aspect
information the intensity (line or continuum) is gridded into a two
dimensional matrix with pixels of size $0$\farcm $4 \times 0$\farcm $4$.
The sky matrix thus derived is a convolution of the intensity distribution
of the source and the bipolar (due to sky chopping) PSF. Using the PSF
determined from Jupiter observations the data  was deconvolved using a
Maximum Entropy Method (MEM) based deconvolution scheme \citep{ghosh88} to
obtain the intensity distributions of the [CII] line and the continuum. The
deconvolved maps of the [CII] line (and continuum) intensity distributions
obtained from the two independent datasets (Up and Down) match well. This
confirms the reliability of most of the observed features. The final [CII]
line and continuum maps are obtained by averaging the respective Up and
Down datasets.

\section{Observational Results}
\label{orion}
\subsection{The [CII] map}

\begin{figure*}
\begin{center}
\includegraphics[angle=0,width=14.0cm,angle=0]{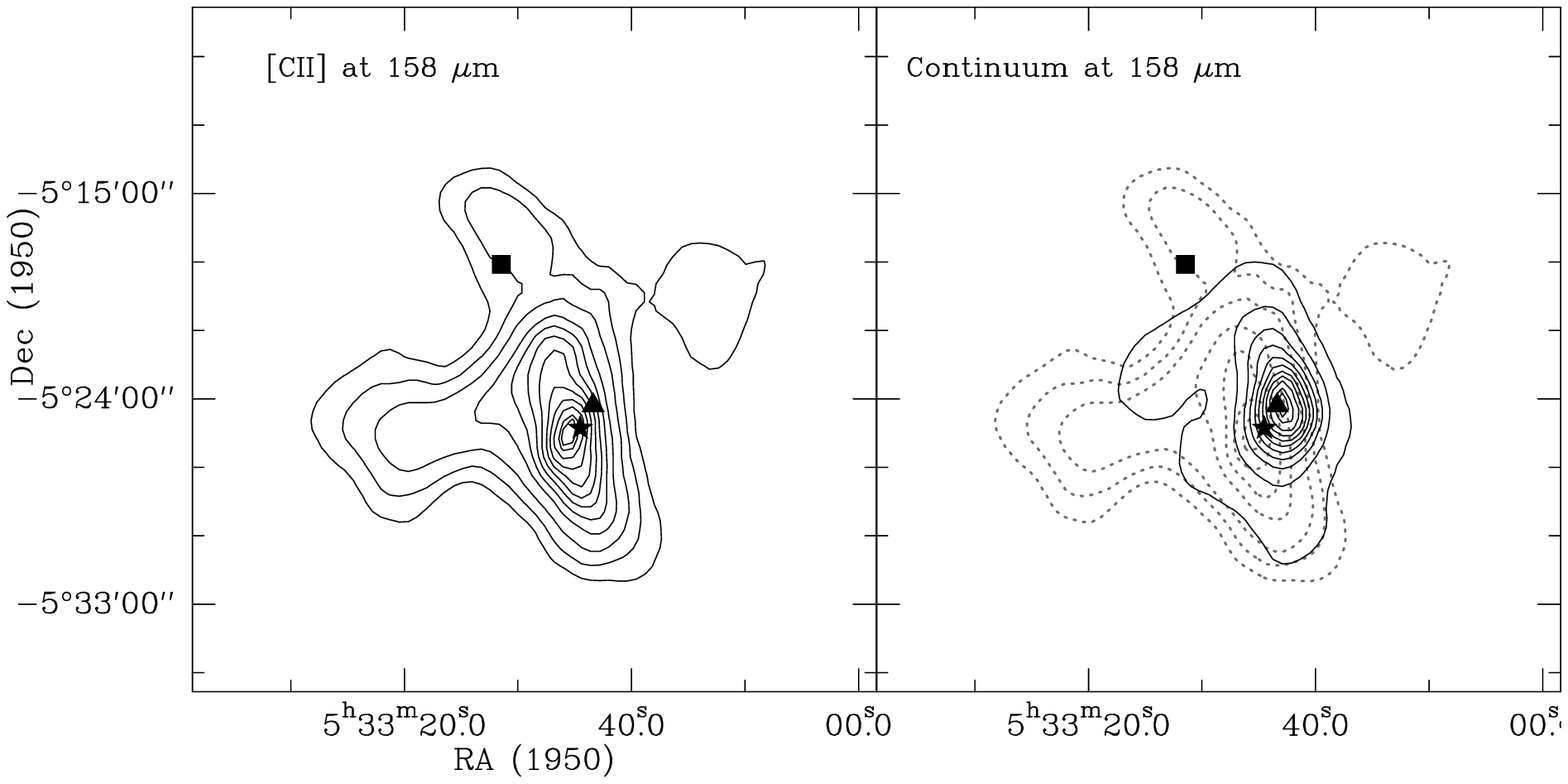}
\caption
{Intensity maps of: $Left:$ [CII] at $158$~\micron\ and $Right:$dust
continuum (overlayed with the [CII] map at $158$~\micron\ contours in
dotted lines). Peak of the [CII] intensity map is
$3.9\times10^{-3}$~erg~s$^{-1}$~cm$^{-2}$~sr$^{-1}$ and that of the
continuum is $4.76\times10^{-3}$~erg~s$^{-1}$~cm$^{-2}$~sr$^{-1}$. The
contour levels are $95\%$, $90\%$ to $10\%$ (in steps of $10\%$) $5\%$
and $2.5\%$ of the respective peaks. Marked with the filled triangle is
BN/KL ( $05^h$ $32^{m}$ $46$\fs$7$ -$05^{\circ}$
$24$\arcmin\ $17$\arcsec (1950)), $\ast$ is  $\theta^{1}$ Ori C (
	 $05^h$ $32^{m}$ $49$\fs$0$ -$05^{\circ}$
	$25$\arcmin\ $16$\arcsec (1950)) and the filled square is M
	43 (  $05^h$ $33^{m}$ $03$\fs$0$ -$05^{\circ}$
	$18$\arcmin\ $06$\arcsec (1950)).
\label{orimap}}
\end{center}
\end{figure*}

Fig.~\ref{orimap} shows the fully sampled maps of velocity-integrated
[CII] intensity and the continuum emission at $158$~\micron. The [CII]
emission peaks at the position of $\theta^{1}$ Ori C. Features detected
in the [CII] emission are similar to those observed by \citet{stacey93}
and \citet{herrmann97}. Both of these previous observations done with an
effective beamsize of $55$\arcsec\ were confined (particularly along the
right ascension) to the immediate vicinity of the central part of
Orion~A, while \citet{herrmann97} extended the map northward along the
ridge. The present observations, though with slightly lower angular
resolution cover a larger extent and detect a strong emission far
towards the east as well as from M43 in the north. The [CII] map also
exhibits an additional emission peak to the west of the main structure
in the Orion~A region. The extended nature of the [CII] emission
supports the present understanding that the [CII] emission not only
arises from the direct interface between the ionized region (produced by
the $\theta^{1}$ Ori C) and the molecular cloud, but also from neutral
atomic interfaces deeper into the molecular cloud produced by the
diffuse UV radiation field. There is no enhancement of [CII] emission
close to the outflow associated with BN/KL; this is consistent with the
theoretical models \citep{hollenbach89} which predict that not much
[CII] is produced in shocks. The [CII] emission falls off sharply
towards the west, but appears to be more extended in the east. 
 This conforms well with the currently understood structure of the
region \citep{wilson97}. This model assumes that the dense gas near
the ionizing star shields the more distant gas to the east from the flux
of the ionizing photons and to the west there being less dense neutral
matter, the ionized gas flows in and dominates. The separated emission
peak towards the west has also been detected in
[CI](1-0) and CO(3-2) emissions, as will be discussed in detail later.
[CII] emission detected from the HII region M43, possibly arises from
the
associated HII region which is energized by the star NU Ori.

\subsection{Comparison of  [CII] \& Continuum emission at $158$~\micron}

Comparison of the [CII] emission with the continuum emission at
$158$~\micron\ (right panel of Fig.~\ref{orimap}) reveals that although
there is an overall correlation between the two, there are distinct
differences as well. The salient difference in features between the
continuum and the [CII] emission are: (1) the positions of the peaks of
the two emissions are different and (2) the [CII] emission is more
extended than the continuum emission.

In contrast to the [CII] emission which peaks at the position of
$\theta^{1}$ Ori~C, the continuum emission peaks at the position of the
BN/KL object. This possibly indicates that while the BN/KL object
corresponds to the strongest embedded energizing source, giving rise to
the dust emission peak, it is still not evolved enough to
photodissociate gas which can emit [CII] strongly.   An
alternative or additional explanation for the difference in the peak
positions could be the higher dust density close to the BN/KL object.

The [CII] emission is more extended compared to the continuum emission,
indicating the extended presence of neutral photodissociated gas and
leaking of the UV photons from $\theta^{1}$ Ori~C. The continuum emission
in contrast falls off sharply with increasing distance from the embedded
young stellar objects responsible for  heating the dust in this region.

However we point out that for the lower contour levels, the sensitivity of
the continuum emission presented here is limited. This also accounts for
the lack of the northward and eastward extension of the continuum emission
as compared to the [CII] emission. This becomes apparent if we compare the
features of the continuum map with those in the $138$~\micron\ map by
\citet{bmook00}. While the central features match exceedingly well, the
extended nature of the continuum emission is clearly missed by the present
observations.

\subsection{Gas Heating Efficiency}

The total [CII] luminosity detected in the central region enclosed
within contours up to 20\% of the peak line emission is $45$~\lsun. The
total FIR continuum luminosity was calculated from the flux detected at
$158$~\micron\ up to 6\% of the peak, using the formula proposed by
\citet{thronson86}.   The calculation assumes a dust
temperature of $50$~K \citep{bmook00} and a dust emissivity of
$\lambda^{-1}$ and the calculated FIR continuum luminosity is $1.1\times
10^{5}$\lsun.  The choice of the lowest contour levels in the [CII] and
continuum maps, up to which we integrate ensures integration over
similar regions. Thus the [CII] luminosity detected is $0.04\%$ of the
total FIR continuum luminosity. 

Large scale survey of the Galactic [CII] emission \citep{nakagawa98} using
the  BICE shows that the [CII]/FIR ratio is low towards the major star
forming regions and the ratio is $\sim$ 0.2\% for compact sources. The
ratio presented in \citet{nakagawa98} uses the total continuum emission
derived between 40 and 120 \micron. If the same ratio is calculated over 1
to 500 \micron, for a dust temperature of 50K and an emissivity index of
2.0 the ratio would be 0.1\%. The BICE survey was conducted with an
effective beamsize of 15\arcmin, which is much larger than the resolution
of our observations (1\farcm5). The resulting difference in the beam
filling factors of the two observations can possibly account for this
discrepancy in the estimated ratio of the [CII] and the FIR luminosities.

The reduction in the [CII] to continuum ratio towards regions of high mass star
formation is explained by model calculations. PDR models
\citep{hollenbach91} show that in high density regions with large G$_{0}$,
 the [CII] line is rather saturated due to its lower
excitation energy and lower critical density, while lines such
as the [O I] $63$~\micron\ line become dominant coolants. In addition,
dust grains being positively charged in regions with high G$_{0}$, the
efficiency of photoelectric heating of gas decreases.

\section{Morphological comparison with other observations:}

We present a comparison of the observed [CII] and dust continuum emission
with the \13CO(1-0), [CI](1-0), CO(3-2) and the radio continuum emission
at $1.5$~GHz. The comparison with \13CO(1-0) is aimed at understanding the
overall column density structure of this region, vis-a-vis the presence of
heating sources. The comparison with [CI] and CO(3-2) emissions is aimed
towards understanding the volume density structure and the distribution of
the UV field, in an attempt to derive a coherent picture of PDRs giving
rise to all of these emissions. The comparison of [CII] emission with the
radio continuum emission is intended to identify ionized gas contributing
substantially to the [CII] emission.

\subsection{Comparison with \13CO(1-0): Column density tracer}

\begin{figure}[h]
\begin{center}
\includegraphics[angle=0,width=7.0cm,angle=0]{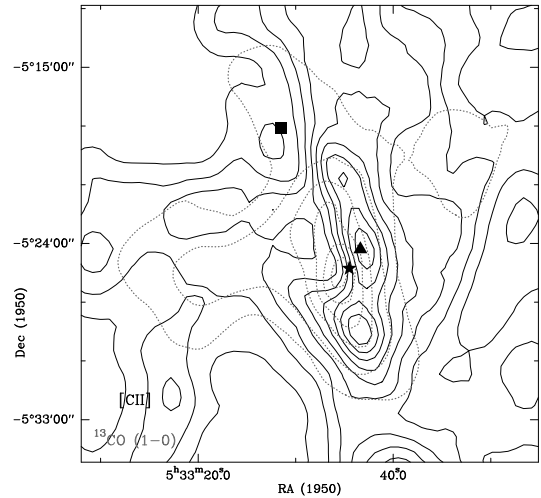}
\caption
{Overlay of the [CII] at $158$~\micron\ (contours as dotted lines) and
\13CO (1-0) \citep{plume00} emissions from Orion A. \13CO (1-0) has been
smoothed to a resolution of $1$\farcm$5$. Peak \13CO (1-0) intensity is
32.7 K km s$^{-1}$, contours levels are 10\% to 90\% in steps of 10\% and
95\% of the peak. {The contour levels for the [CII] map are 2.5 10 30 50
70 90 \% of the peak intensity.} The marked
sources in the map are identical to Fig.~\ref{orimap}.
\label{cii13co}}
\end{center}
\end{figure}

Observations suggest that $^{12}$CO is optically thick ~\citep{castets90}
in Orion A. On the other hand, \13CO\ is optically thin in most cases and
is a better tracer of the column density of the neutral molecular gas. We
compare our [CII] data with that of \13CO(1-0) observed with a beam of
$47$\arcsec, at the FCRAO by \citet{plume00}. Fig.~\ref{cii13co} also
shows an overlay of the [CII] emission with the \13CO\ (1-0) emission. The
\13CO\ (1-0) data has been integrated over velocities ranging between $5$
and $12$~km/s, smoothed to the resolution of the [CII] map
($1$\farcm$5$) and regridded on a $0$\farcm$4$ grid. 

The velocity integrated \13CO\ (1-0) intensity map shows two prominent
peaks, one coinciding with the infrared object BN/KL and the other lying
to the south of it. The continuum emission also peaks at the location of
the BN/KL object, thus implying a core of high column density (and
possibly high volume density as well) with an embedded heating object at
a reasonably early stage of evolution. The southward extension of the
dust continuum emission is resolved into a clump in the \13CO(1-0) map.
However, neither the \13CO(1-0) nor the dust continuum map show the
separated emission peak to the west, which is seen in the [CII] map. The
[CII] emission in contrast shows a single peak at the position of the
ionizing star $\theta^{1}$ Ori C. Since both [CII] and \13CO\ (1-0)
emission are to a good extent optically thin, the dissimilarity in the
overall emission features hint at the difference in the phases of the
ISM which dominantly contribute to these emissions.  The \13CO(1-0)
emission originates from cold molecular material and hence is more
widespread. We note that there is a distinct similarity in the eastward
extension of the \13CO(1-0) emission with that of the [CII] emission.
The \13CO(1-0) map shows an isolated peak at the position of M43 to the
north, which is a little offset from the [CII] intensity enhancement in
this direction.

\subsection{Comparison with CI(1-0), CO(3-2): PDR view}

The present understanding of the PDRs in molecular clouds suggests that
the [CII] emission from the PDRs arises from the surface of UV
irradiated clumpy molecular clouds. The penetration of UV into the
molecular material creates a stratification in which several carbon
bearing species appear and disappear sequentially with increasing visual
extinction into the cloud. Other species which are present in the PDRs
and contribute to the cooling of these regions include, the neutral
atomic carbon and the CO (mid-J rotational transitions) from close to
the edge of the UV irradiated surface and  the low-J rotational
transitions of CO and its isotopomers which trace the large scale gas
distribution of the molecular cloud.

\begin{figure*}
\begin{center}
\includegraphics[angle=0,width=14.0cm,angle=0]{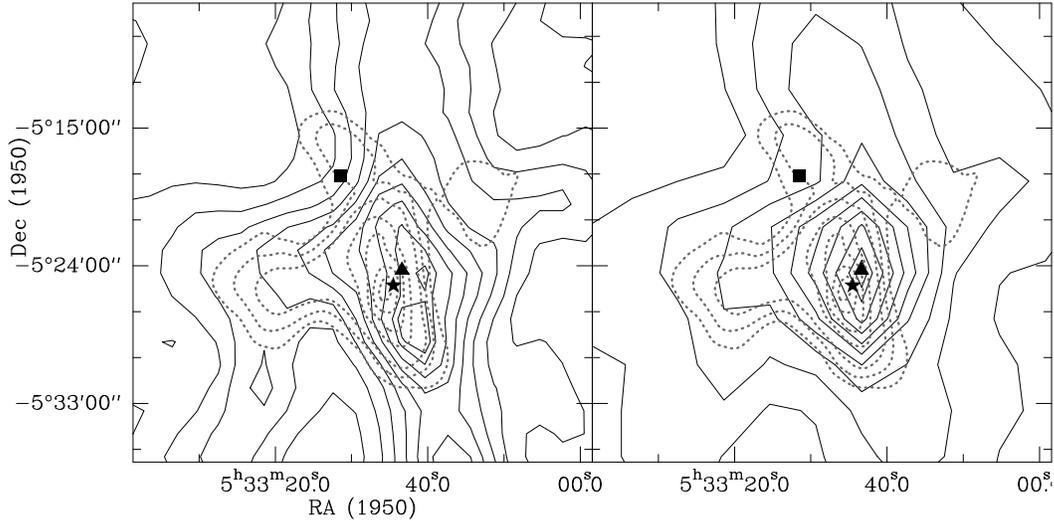}
\caption
{Overlay of [CII] map (with $Left:$ [CI] (1-0) \& $Right:$ CO (3-2) maps
of Orion A \citep[Resolution of $2$\farcm$1$ \& $3$\arcmin\
respectively, CI data smoothed to $3$\arcmin;][]{ikeda99}. Peaks of [CI]  \& CO(3-2) are 41.4 K km
s$^{-1}$  and 382.4 K km s$^{-1}$ respectively. Contour levels for both
maps are 10\% to 90\% in steps of 10\% and 95\% of the peak. Sources
marked are the same as in Fig.~\ref{orimap}. Both maps are at an angular
resolution of $3$\arcmin\ and gridded on $1$\farcm$5$ grid. Relevant
details for the [CII] map as well as the marked sources in the map are
identical to Fig.~\ref{cii13co}.
\label{ci_co}}
\end{center}
\end{figure*}

Fig.~\ref{ci_co} shows the measured  [CI] $^3$P$_{1}$-${^3}$P$_{0}$
($left$) and CO (3-2) ($right$) emission  from the same region of Orion~A,
as we have mapped in [CII].   The [CI] and CO(3-2) data have been taken
from the paper by \citet{ikeda99}.  The [CI] and the CO (3-2) data were
observed with angular resolutions of 2\farcm2 and 3\arcmin\ respectively.
Both datasets were presented in the paper with an angular resolution of
3\arcmin\ and on a $1$\farcm$5$ grid spacing. For proper morphological
comparison, we have smoothed the [CII] data to a resolution of $3$\arcmin.

[CI] arises only from the neutral phase of the ISM, while CO (3-2) traces
the warm and/or dense gas. In other words the CO(3-2) emission stems from
the possible clumpy structures in the ISM which also harbour the [CII] and
[CI] emitting regions closer to their surfaces. The [CI] emission
originating from the neutral gas is more widespread than the [CII]
emission and peaks at two positions, none coinciding either with the BN/KL
object or $\theta^{1}$ Ori C. The southern peak matches with the second
peak of the \13CO(1-0) map. We point out that the peaks of the emissions
of the three lines follow the sequence [CII]/CO/[CI], which is different
from the PDR scenario in which the dominant carbon bearing species changes
from C$^{+}$ to C$^{0}$ to CO. The eastward  extension and the separated
emission peak to the west seen in the [CII] map are clearly seen in the [C
I] map also. We note that [CI] emission shows no enhancement at the
position of M43 contrary to the expectation that presence of [CII]
emission indicates the presence of PDR and hence [CI] emission. This
indicates that the [CII] emission in M43 originates from the associated
HII region, rather than from the neutral PDR.

At this angular resolution the CO(3-2) map shows much less features as
compared to the [CI] map. The emission peaks at a single position centered
on the BN/KL object, much like the continuum emission. The peak position is
identical to the northern peak of the \13CO(1-0) emission. The extension
towards the east and the separated emission peak to the west as seen in the
[CII] map are also clearly distinguishable in the CO(3-2) map. Presence of
these features in the CO(3-2) map suggests enhancement of temperature and
hence presence of embedded sources. This is also substantiated by the
observed [CII] emission.

\subsection{Comparison with $1.5$~GHz radio continuum}

\begin{figure}
\begin{center}
\includegraphics[angle=0,width=8.0cm,angle=0]{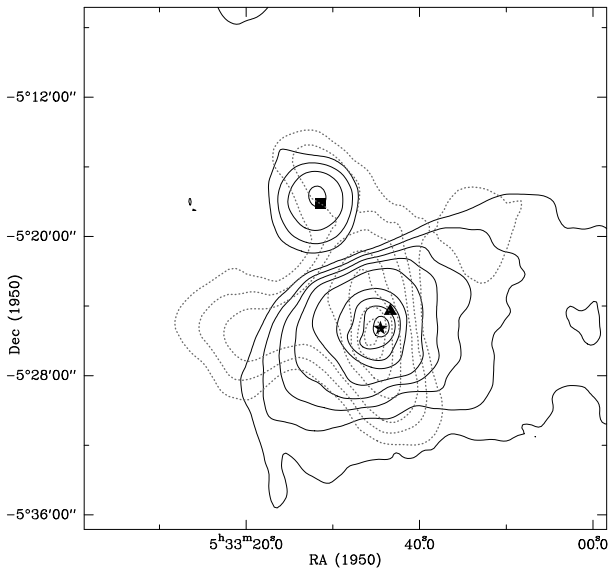}
\caption
{Overlay of VLA $1.5$~GHz continuum image of Orion A \citep{subra01} with
the [CII] intensity map. The peak intensity is 26 Jy beam$^{-1}$ and the
contours are at $0.25$, $1.0$, $2.5$,  $5$, $10$, $30$, $50$, $70$ and
$90$\% of the peak.  Relevant details for the [CII] map
as well as the marked sources in the map are identical to 
Fig.~\ref{orimap}
\label{cii_radio}}
\vspace*{-0.5cm}
\end{center}
\end{figure}

The first ionization potential of carbon is $11.26$~eV (less than that
of hydrogen). Hence, the [CII] emission stems not only from the neutral
PDRs but also from the ionized gas in the HII regions. Model
calculations of dust free HII regions imply that stars hotter than
spectral type O7 produce sufficiently high flux of photons with energies
greater than 24.587 eV to maintain all of the helium in the form of
He$^+$, and therefore all of the carbon in the form of C$^{++}$ (the
ionization energy required to produce C$^{++}$ lies only 0.204 eV below
that required to ionize He). However, presence of dust within the HII
region causes the stellar Ly continuum radiation field to become
depleted of He-ionizing photons. As a result some C$^{+}$ may be found
in dusty HII regions ionized by stars hotter than O7. Alternatively, the
HII regions whose ionizing stars are O8 or cooler are incapable of
maintaining a significant fraction of carbon in the C$^{++}$ state and
hence may have a significant population of \cplus.

Fig.~\ref{cii_radio} shows an overlay of the $1.5$~GHz radio continuum
emission measured using VLA, with an angular resolution of $1$\arcmin\
by \citet{subra01}, with the [CII] map. The basic features of the radio
continuum emission are very different from the
[CII] emission, implying origin in different components (ionized and/or
neutral) of the ISM. The [CII] emission is extended along the
north-south direction, while the radio continuum emission is extended in
the east-west direction. The [CII] halo is more extended toward the
east, while the radio continuum is extended mostly toward the west.
This agrees well with the recent models for the Orion A region
\citep{wilson97}.  The radio continuum emission peaks at the positions
of the two HII regions associated with M42 and M43. The [CII] map peaks
at the position of M42 and shows enhanced emission close to M43, with no
corresponding enhancement in [CI] emission.

We note, M42 is ionized by a O6-O7 star, so that the calculated
[CII] intensity originating in the ionized gas is only 5\%
\citep{russell80}. However M43 is ionized by a B0.5 V star, so the
enhancement of [CII] emission with no corresponding [CI] emission peak
can be attributed to the ionized gas in the HII region.

\section{Discussion}
\subsection{Line Intensity Profile \& Ratios}

In order to probe the density and the UV field strength distribution with
increasing distance from $\theta^{1}$~Ori~C, we derive profiles of the
[CII], [CI], CO(3-2) line intensities for a east-west cut at the
declination of the $\theta^{1}$ Ori C and calculate the intensity ratios
at a few selected points. All three datasets have/are smoothed to an
angular resolution of $3$\arcmin\ and are gridded on a $1$\farcm$5$ grid. 

\begin{figure*}
\begin{center}
\includegraphics[angle=0,width=9.0cm,angle=0]{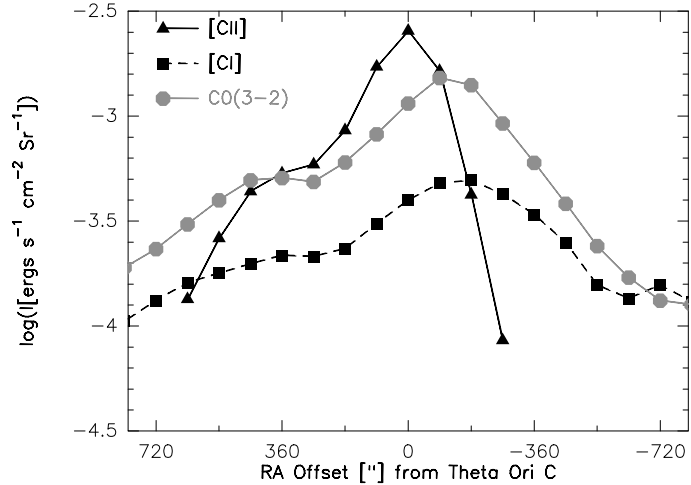}
\caption
{East-west profile of intensities of [CII], [CI] and CO(3-2) at the declination of $\theta^{1}$ Ori~C. The [CI] and CO(3-2) intensities have been multiplied by a factor of $100$.
\label{intprof}}
\end{center}
\end{figure*}

Fig.~\ref{intprof} shows the intensity profiles referred to above, of
[CII], [CI] and CO(3-2) along the east-west direction at the declination of
$\theta^{1}$ Ori C. The [CII] emission falls off rapidly (steeper towards
the west) away from $\theta^{1}$ Ori C, but continues showing residual
emission well beyond. The [CI] emission arising from cold neutral medium
is more widespread. The CO(3-2) emission tracing warmer gas is moderately
distributed, and it's profile shows a hump towards the east which is
similar to that seen in the [CII] intensity profile.


Table~\ref{ratiotab} presents the absolute [CII] intensities, the
C$^{+}$ column densities and ratios observed at several selected
positions (offsets relative to $\theta^{1}$ Ori C) along the E-W cut
shown in Fig.~\ref{intprof} as well as the position of the BN/KL
object (-90\arcsec,90\arcsec).  In order to obtain a first
order estimate the C$^{+}$ column densities have been calculated
assuming the line to be optically thin, the densities to be high enough
so that the emission is thermalized to a temperature of $165$~K
\citep{stacey93} and the beam filling factor to be unity. The C$^{+}$
column densities are thus lower limits. In Table~\ref{ratiotab} the UV
flux and density correspond to the results of the PDR model analysis of
these intensities.  We discuss the PDR modeling in details in the next
subsection.

\begin{table*}
\caption{Ratios of observed line intensities (in K km s$^{-1}$) and the 
modeled parameters, at selected positions.}
\begin{tabular}{cccccccccc}
\hline\hline
\multicolumn{1}{c}{Position$^{1}$}&
\multicolumn{1}{c}{I$_{[CII]}$}&
\multicolumn{1}{c}{N$_{C^+}$}&
\multicolumn{1}{c}{$\frac{\rm [CII]}{\rm [CI]}$}&
\multicolumn{1}{c}{$\frac{\rm [CII]}{\rm CO(3-2})$}&
\multicolumn{1}{c}{$\frac{\rm [CI]}{\rm CO(3-2)}$}&
\multicolumn{1}{c}{Density}&
\multicolumn{1}{c}{UV flux}&
\multicolumn{1}{c}{$\chi^{2}$}&
\multicolumn{1}{c}{Remarks}\\
&ergs s$^{-1}$ cm$^{-2}$ sr$^{-1}$ & cm$^{-2}$ &&&&log(cm$^{-3}$) &log(G$_0$)&\\
\hline
(-90\arcsec,90\arcsec)  & 1.30(-3)  & 7.3 (17) & 267 & $~78$ & 0.29 & 5.0 & 4.25 & 1.0 & CO(3-2) peak\\
(0\arcsec,0\arcsec)  & 2.55(-3) &  1.4 (18)&  631 & 220 & 0.35 &5.25 & 5.5$\ast$ & \ldots & [CII] peak\\
(180\arcsec,0\arcsec) & 8.57(-4)  & 4.8 (17) & 361 & 141 & 0.39 & 5.0 & 4.75 &0.7\\
(-180\arcsec,0\arcsec)  & 4.21(-4)  & 2.4 (17) & $~84$ & $~30$ & 0.35 & 4.75 & 2.5 & 1.6 &\\
(360\arcsec,0\arcsec) & 5.30(-4)  & 3.0 (17)  & 239 & 104 & 0.43 & 4.75 & 3.75 & 0.6 & \\
\hline
\hline
\end{tabular}

{\small 1 Offsets relative to the position of $\theta^{1}$ Ori C
($\alpha_{1950}$ = $05^{\rm h}$$32^{\rm m}$$49$\fs$0$, $\delta_{1950}$ =
-$5$\degr $25$\arcmin $16$\arcsec)}

{\small $\ast$ Since the PDR model gives no solution, the UV flux has been
calculated using the luminosity of Trapezium cluster (L$\sim 10^{5}$\lsun)
and the distance between the Trapezium cluster and the molecular cloud (d
= 0.25pc \citet{odell01})}
\label{ratiotab}
\end{table*}

\subsection{Density and UV Flux from PDR Model}

We have used the PDR model of \citet{kaufman99} to interpret the
observed intensity ratios of [CII], [CI] and CO(3-2) lines and to
estimate the volume density and the UV field strength at selected
positions.  Table~\ref{ratiotab} presents the observed line ratios and
results of modeling including the values of the reduced $\chi^{2}$ at
each of these positions. The models by \citet{kaufman99} are similar in
terms of chemistry, radiative transfer, thermal balance etc. to those
presented by \citet{tielens85}, \citet{hollenbach91} and
\citet{wolfire90}. But they use updated reaction rates, heating rates as
well as chemical abundances and most importantly they take into account
the heating due to PAHs which is important for the interpretation of the
FIR cooling lines like [CII] and
[OI]. \citet{kaufman99} have created a database of PDR models, for a
density range from n$=10$ to $10^{7}$~cm$^{-3}$ and for a
far-ultraviolet (FUV) radiation field ranging between G$_{0}=0.3$ to
$3\times 10^{6}$ (where G$_{0}$ is in units of  an average
interstellar flux of $1.6\times10^{-3}$~ ergs s$^{-1}$ cm$^{-2}$).

\begin{figure*}
\begin{center}
\includegraphics[angle=0,width=10.0cm,angle=0]{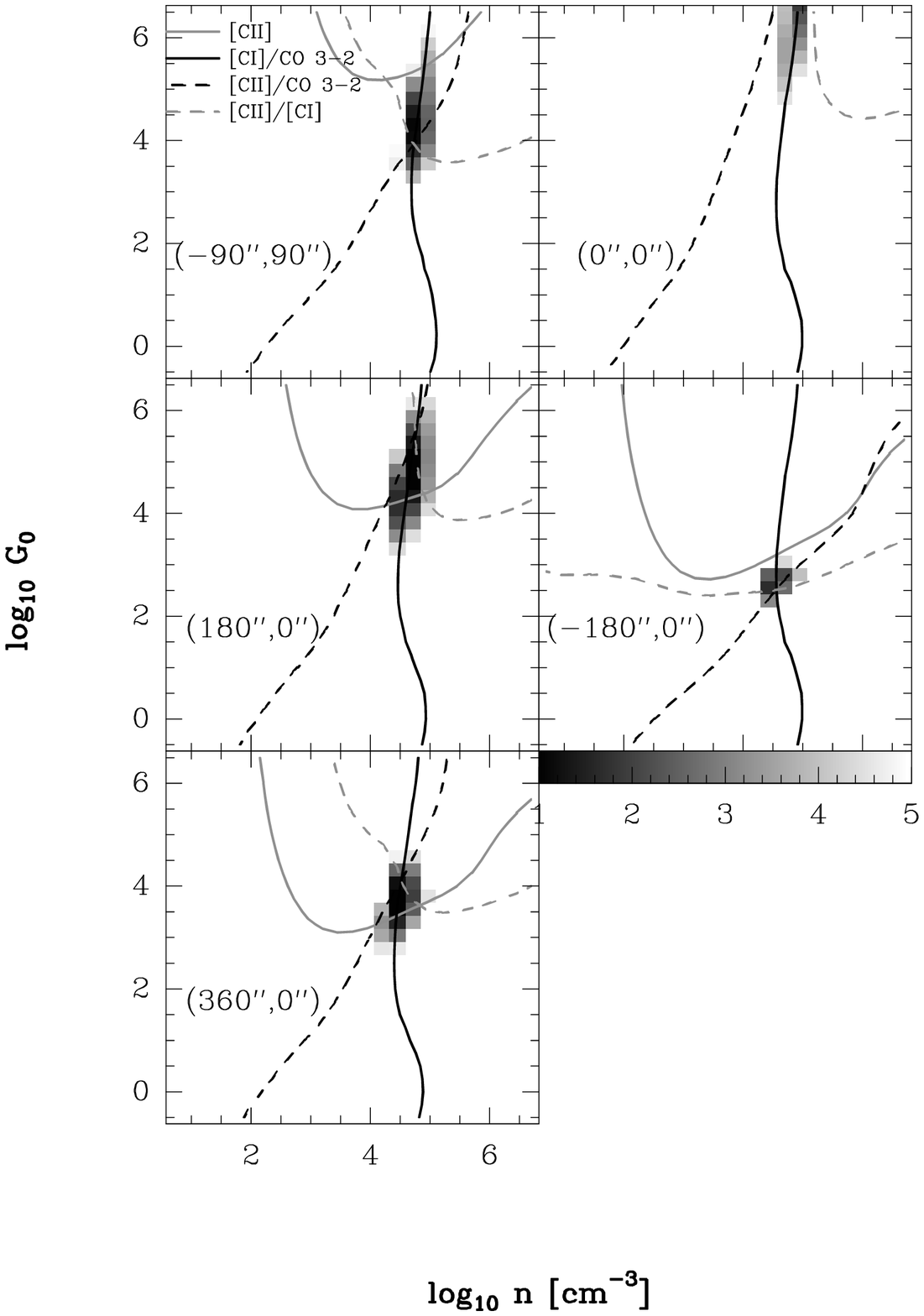}
\caption
{Each panel corresponds to a selected position, the offsets of which are
expressed relative to the position of $\theta^{1}$ Ori C. Plotted in grey
scale is the reduced-$\chi^{2}$ (absolute values shown in the wedge)
calculated from a comparison of the predicted intensity ratios and [CII]
intensities by the Kaufman model, over the entire parameter space of
volume density (in units of cm$^{-3}$ and UV flux (in units of Habing
field). The darker regions correspond to lower values of the
reduced-$\chi^{2}$. The details of the drawn lines are given in the top
left panel and are identical for all panels.
\label{modelplot}}
\end{center}
\end{figure*}

In order to determine the densities and UV fluxes which can explain
intensity ratios, we have done a $\chi^{2}$ fitting of the observed
ratios with respect to the model predictions. Fig.~\ref{modelplot}
shows in grey scales the reduced $\chi^{2}$ for the observed line ratios
at the selected positions.  The grey scale is proportional
to the $\chi^{2}$ and weighted with the individual observational
uncertainties. Based on the estimated calibration errors of the
individual observations, we take an uncertainty of $30$\% for the ratios
involving [CII], while for the
[CI]/CO(3-2) ratio the uncertainty is taken to be $20$\%. The lines
correspond to the observed intensity ratios and the absolute [CII]
intensity (as described in the legends). The values for the volume
density and UV flux at each position were determined from the location
of the minimum $\chi^{2}$. We discuss the result of modeling for each
position separately.

\subsubsection{Positions away from the [CII] peak}

{\em (-90\arcsec,90\arcsec):} This position corresponds to the peak of the
CO~(3-2) map and is approximately the location of the BN/KL object. The best
fit model corresponds to an UV field of G$_{0} = 10^{4.25}$ and a
volume density of $10^{5}$ cm$^{-3}$. The density compares well with
those derived from the multi-line study of Orion A by \citet{wilson01}.
We point out that the fitted model is not accurate; though it explains
the line intensity ratios, the derived physical conditions predict
an absolute [CII] intensity lower than the observed value.

{\em East of $\theta^{1}$~Ori~C:} We are able to explain all the
intensity ratios as well as the [CII] intensity at the $2$ eastern
positions ($180$\arcsec,$0$\arcsec) and ($360$\arcsec,$0$\arcsec) using
the PDR models. 

{\em West of $\theta^{1}$~Ori~C:} We restrict the analysis of the western
side to a single point (-$180$\arcsec,$0$\arcsec) owing to the limited
dynamic range of the [CII] map. The model fitted is acceptable, although
we note that the predicted [CII] intensity is lower than the observed
intensity.

\subsubsection{The [CII] peak}

{\em (0\arcsec,0\arcsec):} This corresponds to the position of the star
$\theta^{1}$ Ori~C. For the parameter space explored the models do not have
any solution. We identify the failure as due to the excess of observed
[CII] emission over the model predictions. We note that if models
corresponding to higher UV field were considered,  a solution could be
derived, but it would not have been realistic. This is because for the
given luminosity ($\sim 10^{5}$\lsun)  of the ionizing stars, the Trapezium
cluster, and their distance of 0.25~pc from the molecular cloud
\citep{odell01} the strength of the UV field at this position would at 
most be $3.3\times 10^{5}$G$_{0}$. 

Since the ionizing star is of the type O6 the contribution of the
ionized gas towards the [CII] emission will be negligible, unless the
high dust density plays a crucial role. However this scenario for the
present is highly speculative, so we do not persist with it. Instead we
keep in mind that the geometry and structure of Orion A is far too
complicated and not fully understood to be satisfactorily modeled by the
presently available PDR models. We consider the effect of inappropriate
model geometry  as a probable reason behind the failure of the PDR
models to generate a satisfactory solution.

We first estimate the C$^{+}$ column density which is needed to explain
the detected [CII] intensity, assuming that the line is optically thin.
For regions outside HII regions the almost temperature independent,
critical density (n$_{\rm crit}$) for collisional excitation is $4\times
10^{3}$ cm$^{-3}$. Since the measured volume densities for these regions
are well in excess of these values (typically $\sim 10^{5}$ cm$^{-3}$),
the [CII] emission is in thermal equilibrium. \citet{stacey93} suggest
that the minimum gas kinetic temperature of the region is $\sim 165$~K.
With these assumptions and values we estimate the C$^{+}$ column density
to be $1.4\times 10^{18}$ cm$^{-2}$. Assuming an abundance ratio of
[C$^{+}$]/[H] $\sim 3\times 10^{-4}$, this corresponds to a hydrogen
column density of $4.8\times 10^{21}$ cm$^{-2}$, which is equivalent to an
A$_{\rm v}$ of $2.5$~mag. The model of \citet{kaufman99} treats the PDRs
for A$_{\rm v}$ upto $10$ mag, but being a homogeneous slab model
considers the [CII] emitting region to be confined to the surface.  If
instead of a single such surface, a number of such surfaces (optically
thin for [CII] emission) are placed, then it is possible to enhance the
C$^{+}$ intensity relative to the other line intensities. The need to
invoke a number of such surfaces, suggests the presence of non-homogeneity
or clumpiness in the ISM. The scenario changes appreciably if a spherical
clump model of PDRs is used, in which case the width of the [CII] emitting
layer changes depending on the size of the clump \citep{koster94}. A model
treating an ensemble of spheres is more realistic, since observations
support that the ISM consists of clumps immersed in a tenuous inter-clump
medium, rather than being a homogeneous slab. \citet{mochizuki94} have
explicitly discussed that in the clumped scenario, the [CII] emission is
enhanced significantly compared to the homogeneous case, if the
average size of the clumps are such that A$_{\rm V} \ll 10$~mag.

\subsection{Summary of PDR modeling}

The PDR models are able to explain the observed line intensity ratios,
satisfactorily for positions away from the energizing sources like the
$\theta^{1}$ Ori C and the BN/KL object. At the position of the [CII]
peak in particular it fails to provide a solution, mainly because of the
excess in [CII] emission over the model prediction.  This
is probably because the actual geometry is far more complex than the
model assumptions.  However, in the absence of velocity-resolved [CII]
observations it is difficult to  throw more light  on the
actual structure of the [CII] emitting region. We note that although
the contribution of the ionized gas towards the [CII] emission
 at the peak position in M42 is expected to be negligible it may
not be totally absent and must be taken into account while attempting
better constrained modeling at other positions. 

According to the PDR model, the [CI]/CO(3-2) intensity ratio constrains
the volume density, while the [CII]/[CI] is very sensitive to the
strength of the UV flux. For all the selected positions, the
[CI]/CO(3-2) intensity ratio varies very little and the derived volume
densities are almost the same. The larger variation of [CII]/[CI]
intensity ratios corresponds to a larger spatial variation of the UV
field strength.

\section{Summary}

We have presented the first results of observations with a Fabry Perot
spectrometer tuned to [CII] $157.7409$~\micron\ onboard a balloon-borne
FIR telescope, \fps100. These observations demonstrate the  capabilities
of the instrument we had planned for. The line integrated intensity map of
[CII] emission from the Orion~A region as well as the continuum emission
at $158$~\micron\ have been obtained here. The [CII] luminosity is $0.04$
\% of the FIR continuum luminosity. Morphological comparison of the [CII]
map with with [CI] at $492$~GHz, CO (3-2) and \13CO\ (1-0) data show (i)
that the peaks of the emission occur in the sequence of [CII]/CO/[CI]
which is contrary to the PDR scenario and (ii) the [CII] emission from M43
in the absence of corresponding [CI] emission stems from the HII region
and not from the PDR.

The [CII], [CI] and CO (3-2) line intensity ratios have been modeled
using the PDR models of \citet{kaufman99}. The models though somewhat
simplistic, considering homogeneous plane parallel slabs instead of the
clumpy structure seen observationally, are a first step towards
understanding the observations. The models do not reproduce the observed
[CII] intensity as well as the intensity ratios involving [CII] for
positions which are close to the ionizing source (viz., $\theta^{1}$ Ori
C and possibly BN/KL). This possibly is  an effect of the
assumption of plane parallel geometry instead of the more complicated
structure of the region. The models produce acceptable solutions for
regions with diffuse UV radiation. Similar to the [CII] intensity
profile both the incident UV flux and the volume density show sharper
decline towards the west of $\theta^{1}$ Ori~C, than towards the east.
For all the selected positions the volume density derived using the PDR
models varies little, while the UV field strength varies substantially.

\acknowledgements 

We thank the referee, T. L. Wilson for his comments which have improved
the scientific contents of this paper.  It is a pleasure to thank all
members of the Infrared Astronomy Group of TIFR and the Balloon Group
and Control Instrumentation Group of the TIFR Balloon Facility,
Hyderabad, for their support during the balloon flight.  We thank M.
Ikeda for providing the [CI] and CO (3-2) datasets,  F. Bensch for
providing the \13CO\ (1-0) dataset and R. Subrahmanyan for VLA $1.5$~GHz
continuum image. We also thank M. Wolfire for generating the ratio plots
for the PDR models specially for this work. BM acknowledges the research
fellowship of the Alexander von Humboldt Foundation, Germany.  This
research has made use of the SIMBAD database, operated at CDS,
Strasbourg, France.

%

\end{document}